\documentclass[journal]{IEEEtran}
\ifCLASSINFOpdf
  \usepackage[pdftex]{graphicx}
\else
\fi
\usepackage[tight,footnotesize]{subfigure}
\newcommand{\CO}{$ CO_2 $}

\begin{document}
%
\title{Studies of the Effects of Oxygen and CO$_2$ Contamination of the Neon Gas Radiator on the Performance of the NA62 RICH Detector}
%
%
%

\author{Evelina~Gersabeck
\thanks{Manuscript received November 15, 2011.}
\thanks{Evelina Gersabeck is at INFN, Sezione di Perugia, Perugia, Italy. Now at Physikalishes Institut, Ruprecht-Karls Universitaet, Heidelberg, Germany.}~on behalf of the NA62 RICH Working Group
}

\maketitle

\begin{abstract}
The NA62 RICH detector is used for the separation of pions and muons in the momentum range 15 -- 35 GeV/c and is expected to provide a muon suppression factor better than $10^{-2}$. 

A prototype of the final detector equipped with about 400 PMs (RICH-400 prototype) was built and tested in a dedicated run in 2009.  The $\pi-\mu$ separation was tested, as well as the effect of the pollution of the neon radiator with different amounts of oxygen and \CO. The $\mu$ misidentification probability is about 0.7\% and the time resolution better than 100 ps in the whole momentum range. 

 We did not observe any absorption of the light due to the pollution of the radiator, however an effect on the ring radius is clearly observed due to the change of the change of the refractive index of the medium. The conclusion of the studies is that the amount of \CO~in the final detector should be well known or the quality of the pion identification could be seriously compromised.
\end{abstract}



%
\IEEEpeerreviewmaketitle

\section{Introduction}

%
%
%
%
\IEEEPARstart{T}{he} CERN NA62 experiment~\cite{na62} aims to measure the ultra-rare charged kaon decay $K \rightarrow\pi \nu \bar{\nu}$
(branching fraction $O(10^{-10}))$~\cite{pinunu} with a 10\% accuracy. The detector should efficiently suppress all the backgrounds with a signature similar to the one of the main decay channel. One of the main backgrounds to be suppressed is coming from the abundant $K\rightarrow\mu\nu$ decay.
The main task of the NA62 RICH detector is to separate pions from muons in the momentum range 15--35 GeV/c with a $\mu$-misidentification factor better than $10^{-2}$. In addition to that, it should provide the crossing time of the pion with a resolution better than 100 ps, and an input to Level 0 and Level 1 of the trigger. The important information given by the RICH is the number of the PhotoMultiplier (PM) hits, the ring radius, the center of the ring, and the time of the charged particle.

The test resolution, the light collection technique and, even if biased by the small number of PMs, the Cherenkov angle and the track angular resolutions were checked in a test beam in 2007~\cite{test07}. 
The pion--muon separation capability of the RICH detector prototype equipped with about 400 PMs (RICH-400 prototype) was tested in a dedicated run in 2009~\cite{rich2009}. The prototype was built as a  long tube (about 18 m, and a diameter of about 60 cm) filled with neon (Ne) at atmospheric pressure and room temperature, equipped with a spherical mirror (17 m focal length) at the downstream end and with 414 PMs at the upstream end.  The detector response was tested using a positive hadron beam of variable momentum (in the range 10-75 GeV/c), with a spread of about 1\% and a divergence of about 30 $\mu rad$. At the prototype position it was composed of $K^+$, $\pi^+$, p, $e^+$ with fractions changing with the selected momentum, and a small amount of $\mu^+$ from $\pi^+$ decays. Two different mirrors were tested, both produced by the Marcon company\footnote{MARCON Costruzioni Ottico Meccaniche, via Isonzo 4 - 30027 San Dona' di Piave (VE) Italy (www.marcontelescopes.com).}. One of the mirrors was aluminized and coated with MgF2 by the supplier in 2008, while the latter was aluminized and coated with SiO2 and HfO2 at CERN in 2009 in an attempt to improve the total mirror reflectivity. A small difference of  about 1\% in favour of the new mirror in the number of hit PMs was observed.

The RICH operational range starts above a wavelength of 190 nm and above this value no visible absorption of the light is expected due to the contaminants. However, the impurities change the refractive index of the radiator which could affect the Cherenkov angle and its resolution. Some tests with oxygen and \CO~fractions up to 1\% were done to check the performance of the detector.

\section{Studies of the effect of pollution of the Neon on the performance of the detector}

The data were taken at different momenta, with a beam centered on the mirror at a position (x = 0, y = 0).
 During 2009, several runs with polluted neon were taken as a part of the RICH-400 test beam programme. Two of them were taken with an oxygen pollution of 380 ppm and 425 ppm. 
 In these runs, the old mirror produced by Marcon was used. After changing the mirror with a new one, with improved aluminized coating, new tests with polluted Ne were performed. The gas used for the pollution was \CO.
Four runs were taken for both 0.5\% pollution and 1.0\% pollution with \CO. The beam momentum of the charged particle for these special runs was fixed at  20.0, 26.5, 35.0, and 46.3 GeV/c. 

The data were compared to a dedicated MC simulation based on Geant 4. All the known efficiencies were taken into account: Quartz window transparency, PM Quantum efficiency, Mirror reflectivity, Winston cones efficiency. A detailed geometry description is implemented. In the simulation, the RICH vessel is put in air, and the pion or the electron beam at a given momentum is shot about 6 meters before its window. The mirror reflectivity implemented in the simulation is the one of the old mirror. This is an important remark, as the runs with \CO~pollution are taken with the new mirror. 

\begin{figure*}[]
\vspace{2pt}
\includegraphics[width=42pc]{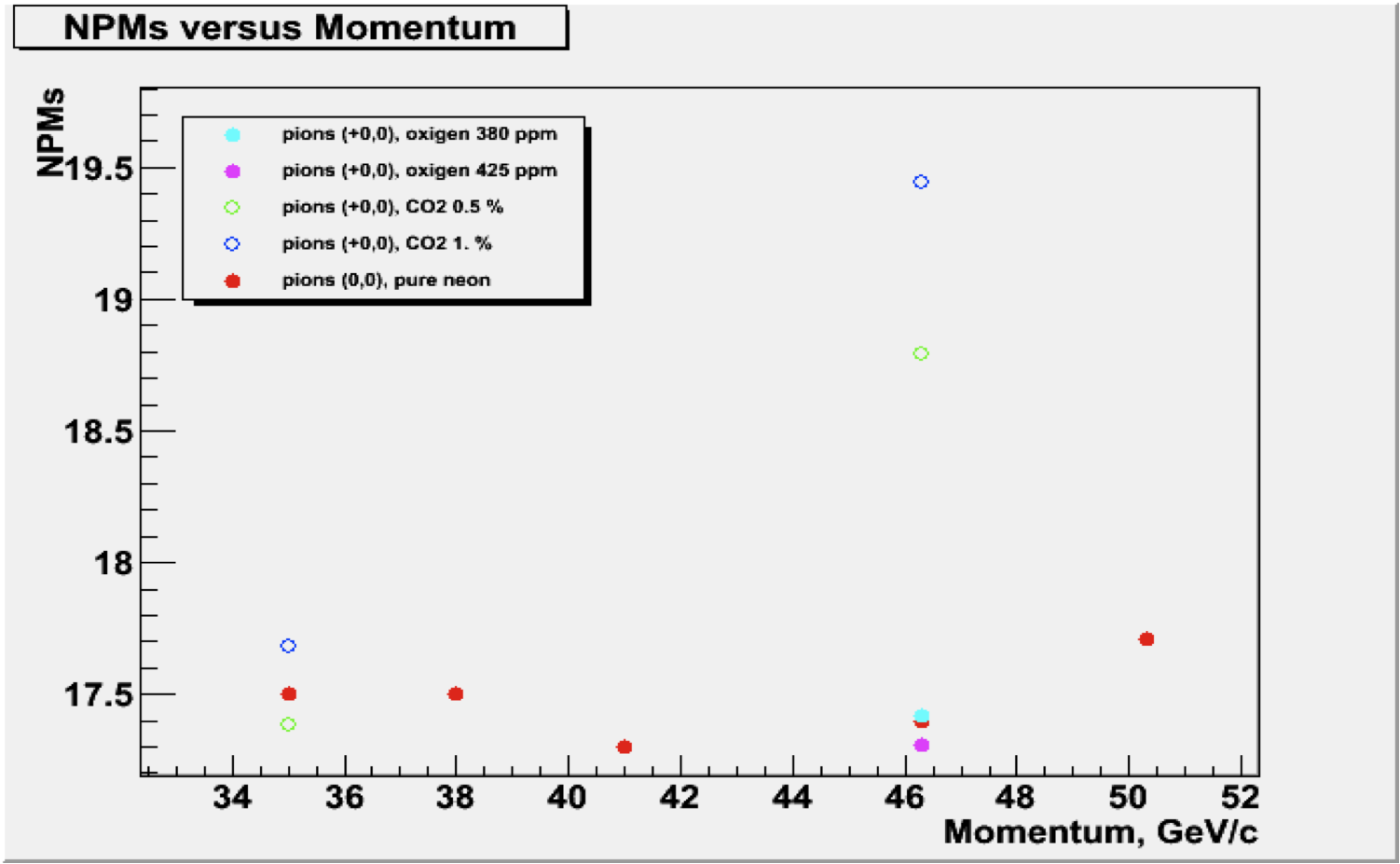}
\caption{Number of hit PMs versus the particle momentum for non-polluted Ne and for oxygen contamination in the Ne radiator. The data with the polluted Ne are in light blue (380 ppm) and purple (425 ppm) dots, hardly distinguishable from the pure Ne data, in red, at 46.3 GeV/c. For \CO~data, plotted with green (0.5\%) and blue (1.0\%) empty circles, the effect on the number of PM hits is visible.}
\label{fig:1}
\end{figure*}

\subsection{Oxygen pollution effects}


The two data runs with oxygen pollution of 380 ppm and 425 ppm were taken at 46.3 GeV/c pion momentum. As it can be seen in Fig.~\ref{fig:1}, these two measurements are hardly distinguishable from the non-polluted data runs. It can be concluded that the oxygen contamination did not affect the number of PM hits, and therefore there is no absorption effect for these amounts of oxygen. 


\subsection{Validation of the Monte Carlo simulation}

\begin{figure*}[]
\vspace{2pt}
\includegraphics[width=42pc]{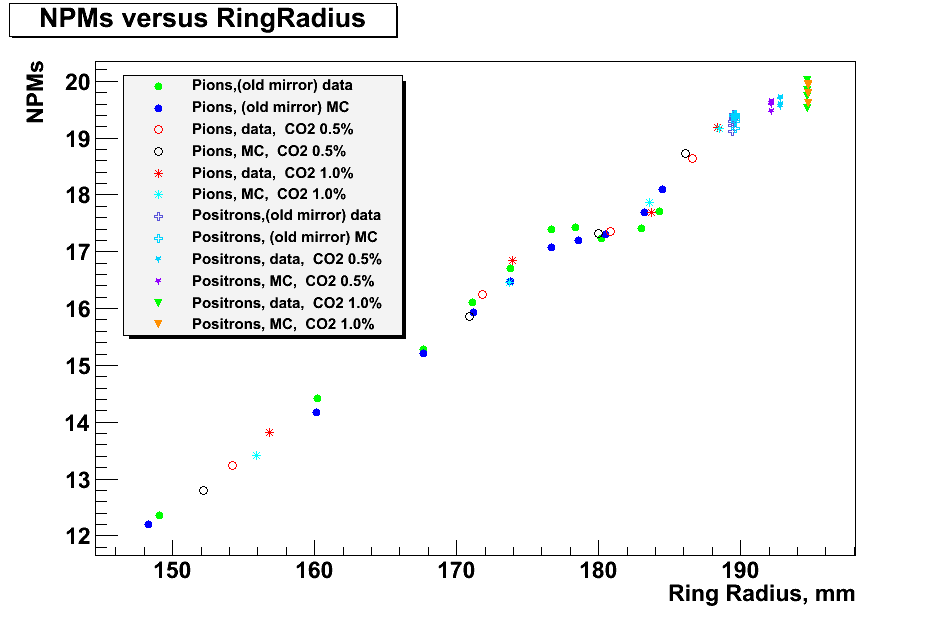}
\caption{Hit multiplicity versus ring radius: comparison of  data and MC simulation for pions and positrons at different momenta. The positron data populates the upper right corner (for positrons $\beta = 1$). A discrepancy in the ring radius for 0.5\% \CO~data is observed (between red and black empty circles for pions, and between light blue and purple stars for positrons).}
\label{fig:4}
\end{figure*}

The data and MC simulation distributions of the ring radius and of the hit multiplicity for non-polluted Ne were compared, in order to validate the MC simulation. In Fig.~\ref{fig:4}, the number of hit PMs is plotted versus the ring radius for pions and electrons at 20.0, 23.4, 26.5, 28.7,31.0, 35.0, 38.0, 41.0, 46.3 am 50.3 GeV/c beam momentum. The runs with Ne contaminated with \CO~were taken at 20.0, 26.5, 38.0 and 46.3 GeV/c. The different momenta correspond to different ring radius for pions (the ring radius increases with the momentum), but for electrons, as $\beta = 1$, the ring radius is always the same\footnote{As for electrons $\beta$ equals 1, the ring radius for electrons is always the maximum possible one, and they are clearly distinguishable.}.

The ring radius is $r = f\Theta_c$, where $f$ is the focal length of the spherical mirror, and $\Theta_c$ is the Cherenkov angle (the angle between the emitted Cherenkov radiation and the particle path). The Cherenkov angle is related to the velocity by $\cos\Theta_c = 1/n\beta$, where $n$ is the refractive index of the medium, and $\beta$ is the velocity of the charged particle.


 \begin{figure*}[]
\vspace{2pt}
\includegraphics[width=42pc]{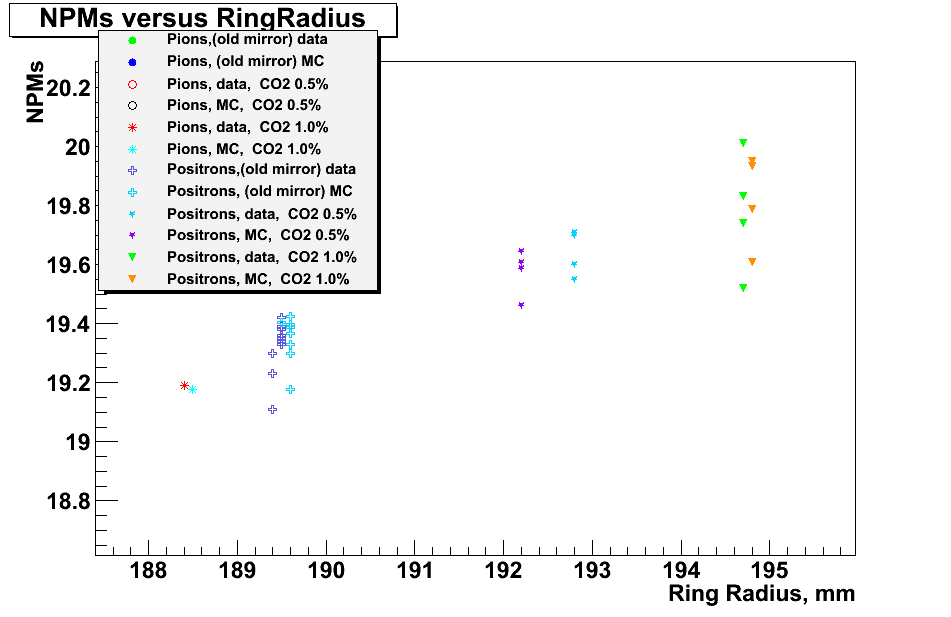}
\caption{Hit multiplicity versus ring radius: comparison of data and MC simulation for pions and positrons at different momenta after scaling factor correction - zoom into the positrons data. The discrepancy between the data for 0.5\% \CO~(light blue and purple stars) is more clearly visible. This figure shows a zoom of Fig.~\ref{fig:4} into the positron data area.}
\label{fig:5}
\end{figure*}

 \subsection{\CO~pollution effects}
 The data taken with and without pollution, together with the MC simulation are plotted in Fig.~\ref{fig:4}. The number of hit PMs increases with the ring radius (and with the momentum). The data with the non-contaminated radiator is used as a reference of the quality of the overall data and MC simulation agreement.
 
 The effect of the pollution on the ring radius is clearly visible: for 1\% of \CO~in the radiator gas the ring radius is 2.7\% larger (measured with electrons).
 
 For the MC simulation of the \CO~ polluted data, a new radiator, with a new refractive index was introduced in the Geant 4 simulation, according to the amount of \CO~used for the different runs. 
 The refractive index $n$ is calculated as
\begin{equation}
n-1=\frac{C}{\nu_0^2-\nu^2}.
\end{equation}

For Ne, $\nu_0^2 = 39160\times 10^{27}$ and $C = 2.61303\times 10^{27}$, while for \CO~these values are $\nu_0^2 = 14097\times 10^{27}$  $C = 6.2144\times 10^{27}$~\cite{bornstein}.
 In case of mixing Ne with \CO, a linear combination was used for the final refractive index
 \begin{equation}
 (n_{mixture}) = (100 - m)\%(n_{Ne})+m\%(n_{CO_2}),
 \end{equation}
 where $m$ is the amount of \CO~ used, and $\nu = 1/\lambda$, where $\lambda$ is the wavelength\footnote{The sensitivity of the photomultipliers chosen (Hamamatsu R7400 U03) is in the range of $\lambda$ from 190 to 620 nm~\cite{lenti}}.

\begin{figure*}[]
\vspace{2pt}

\centering
    \includegraphics[width=7.5cm]{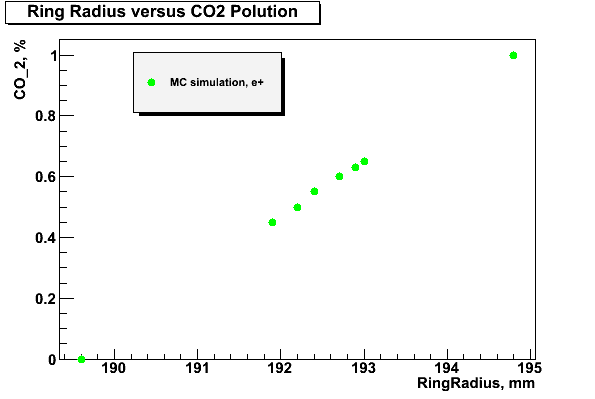} 
    \includegraphics[width=7.5cm]{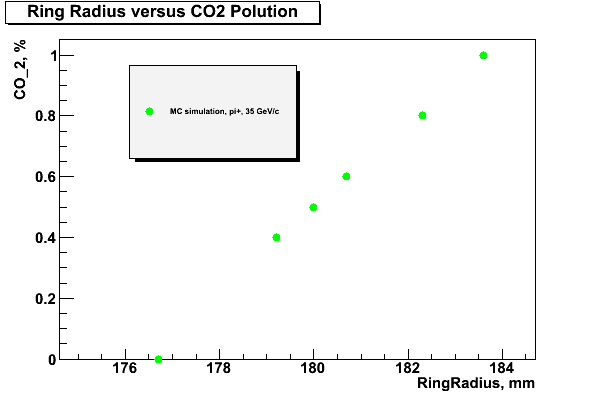} 

\caption{The correlation of the ring radius and the amount of \CO~ for positrons and pions at 35 GeV/c (MC simulation).}
\label{fig:6}
\end{figure*}

 While for data and MC simulation the agreement on the ring radius for the positrons data with no pollution in Ne, and with 1.0\% \CO~pollution, in Fig.~\ref{fig:5} is good, the disagreement on the ring radius for positrons data and MC taken with 0.5\% \CO~pollution is clearly visible. One of the reasons for such disagreement of the ring radii could be a beam with a slightly different momenta than the nominal one. However, this would not give effect for positrons for which $\beta = 1$. From the pion data (Fig.~\ref{fig:4}), the biggest discrepancies for the ring radii of the pions is also found for the 0.5\% \CO~pollution data. In order to explain this mismatch, a hypothesis was made, that the amount of the pollution is different from the measured one.
A test of the correlation of the ring radius and the amount of \CO~in the Ne was made by using MC simulation. The results are shown in Fig.~\ref{fig:6}. The correlation of the \CO~pollution and the ring radius was measured for positrons (see Fig.~\ref{fig:6}, right) and for pions at 35 GeV/c (see Fig.~\ref{fig:6}, left).
This analysis has shown that the discrepancy in the measured ring radius can be explained with a bigger amount of \CO~in the Ne and that it is more likely that the Ne gas is contaminated with 0.6\% \CO. This is compatible with the fact that during the test beam, the amount of \CO~filled in the vessel was controlled by hand as the less significant bit on the apparatus scale was 0.1\%.

\begin{figure*}[]
\vspace{2pt}
\includegraphics[width=42pc]{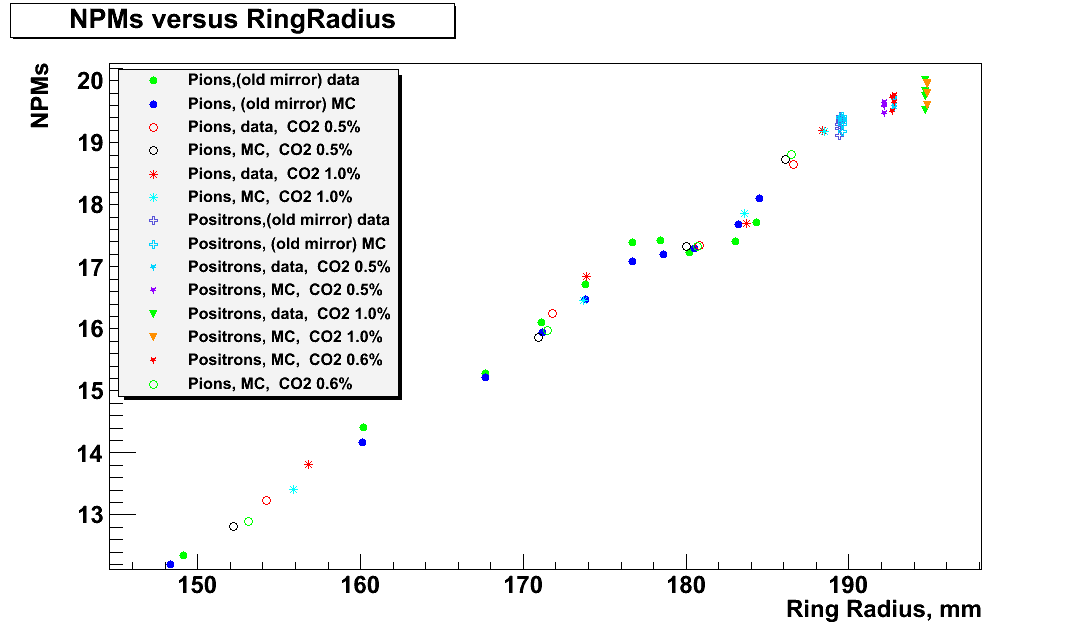}
\caption{Hit multiplicity versus ring radius - data / MC simulation comparison for pions and positrons at different momenta after scaling factor correction. MC simulation of 0.6\% \CO~pollution was added to this plot (empty green circles). The better agreement between the "0.5\%" (empty red circles) data and the 0.6\%  MC simulation (empty green circles) than with 0.5\% MC simulation (empty black circles) is visible.}
\label{fig:7}
\end{figure*}

In Fig.~\ref{fig:7}, MC data for 0.6\% \CO~are plotted, and a better agreement of the test beam data with the simulation is visible.

\begin{figure*}[]
\vspace{2pt}
\includegraphics[width=42pc]{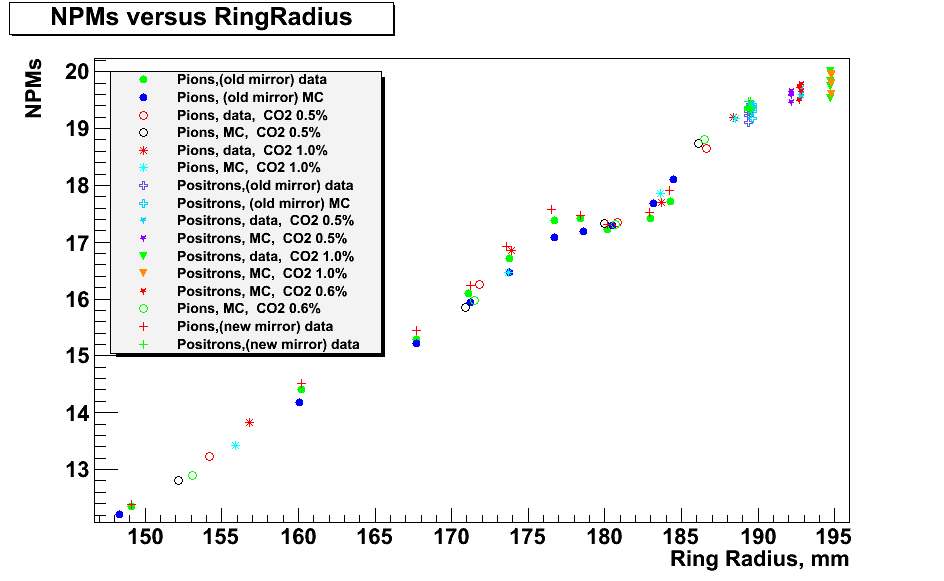}
\caption{Hit multiplicity versus ring radius: comparison of  data and MC simulation for pions and positrons at different momenta after scaling factor correction. The new mirror data for non-polluted runs are added (red (pions) and green (positrons) crosses). Due to the slightly better reflectivity of the new mirror, the reflectivity efficiency is higher, therefore the number of hit PMs is slightly higher (compare green dot with red cross for pions, and dark blue empty cross with red cross for positrons).}
\label{fig:8}
\end{figure*}

As a final step to improve the agreement, the data with the new mirror for the non polluted runs are added to the plot (see Fig.~\ref{fig:8}). As mentioned before, the efficiency of the old mirror is used in the simulation. The comparison between data and MC simulation is done for runs taken with the old mirror. However, the \CO~polluted runs were taken with the new mirror. By adding the non-polluted runs with the new mirror, it is shown that the new mirror has a slightly better reflectivity efficiency. Therefore, if this efficiency is included in the simulation instead the efficiency of the old mirror, we can expect that the number of hit PMs for the polluted data will be slightly increased leading to a better agreement of data and MC for these runs, especially for lower momenta.     

\begin{figure*}[]
\vspace{2pt}
\includegraphics[width=42pc]{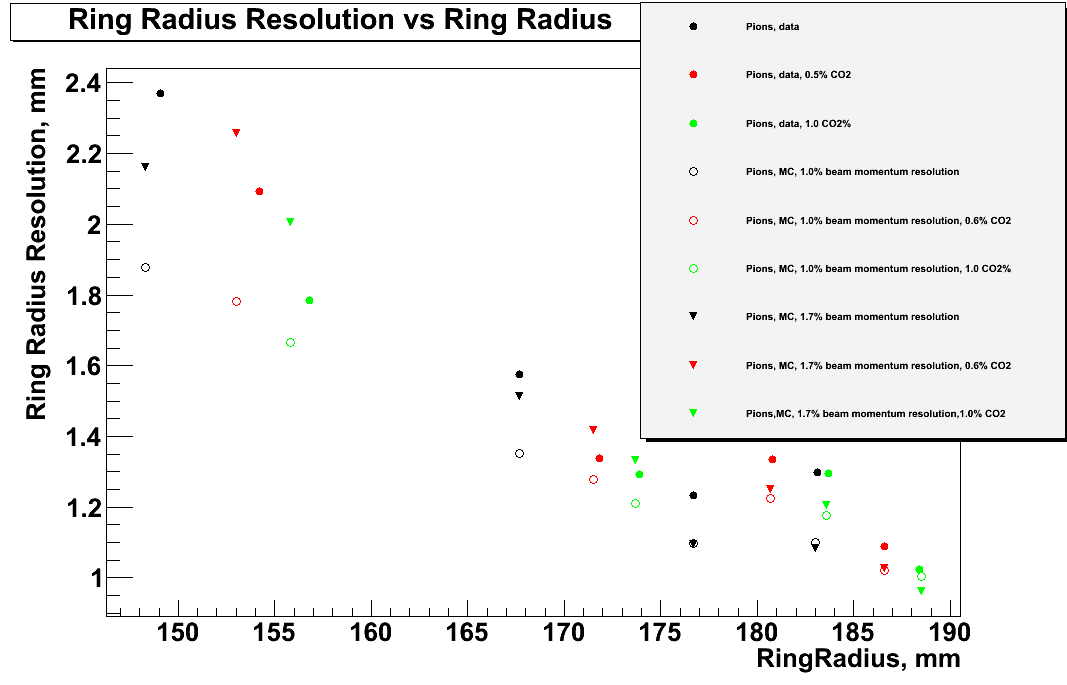}
\caption{The pollution effect on the ring radius resolution. Two sets of MC simulation, with beam momentum resolution of 1.0\% (empty circles) and with 1.7\% (triangles), were compared to the data (dots). }
\label{fig:9}
\end{figure*}

The effect of the pollution on the ring radius resolution is shown in Fig.~\ref{fig:9}.
 The data for 0, 0.5\%, and 1.0\% \CO~are plotted with black, red and green dots, correspondingly.  Then, two sets of MC simulation were added: one generated with beam momentum resolution of 1\% (empty circles with the same color scheme for \CO~pollution as above), and one generated with beam momentum resolution of 1.7\%(black, red and green triangles). The runs used for this study are at 20.0, 26.5, 38.0 and 46.3 GeV/c. 

\section{Conclusions}
No absorption effect on the number of PM hits is observed in several runs taken with polluted Ne gas radiator during the RICH-400 testbeam. However, an effect on the ring radius is clearly observed. For a radiator pollution with 1.0\% of \CO, an effect of 2.7\% on the ring radius is seen. This observation is confirmed by MC simulation. The data which were assumed to be taken with 0.5\% of \CO~are well reproduced by MC simulation with Ne radiator with 0.6\% contamination of \CO. The measurement of the amount of the \CO~could be affected by the limited apparatus precision. The conclusion from these studies is that it is important to know the amount of \CO~in the final detector. In order to easily apply a correction to the ring radius,  $e^+/e^-$ data can be used for additional calibration.  If the amount of \CO~is not known, the quality of the pion identification could be seriously compromised.


%






\ifCLASSOPTIONcaptionsoff
  \newpage
\fi



%

.
%









\end{document}